\documentclass[aps,prb,a4paper,preprint,showpacs]{revtex4}
\usepackage{amsmath}
\usepackage{amsfonts}
\usepackage{amssymb}
\usepackage{slashbox}
\usepackage{graphicx}
\usepackage{amsbsy}

\begin{document}
\title{Conductance in the Helimagnet- and Skyrmion-Lattice-Embedded Electron Waveguide }
\author{Rui Zhu\renewcommand{\thefootnote}{*}\footnote{Corresponding author.
Electronic address:
rzhu@scut.edu.cn}}
\address{Department of Physics, South China University of Technology,
Guangzhou 510641, People's Republic of China }

\begin{abstract}

The helimagnet (HM) and skyrmion lattice (SL) are topologically nontrivial magnetic states. Their spin texture gives rise to finite topological magnetic field and Lorentz force. As a demonstration of the emergent electrodynamics besides the Hall effect, the transmission of electrons within a waveguide (WG) embedded with a HM/SL layer is shaped by the topological spin texture. In this work, we investigated the conductance properties in the HM/SL-film-embedded electron WG and found that under translation of the HM/SL layer the conductance contour as a function of the layer center demonstrates nearly identical pattern to the original spin field contour. When an electron transports in the WG, the topological magnetic field generalized by the HM/SL spin texture exerts on it and distorts the wavefunction shape in the $x$-$y$ plane. By interference between different quantum paths the electron experiences, the HM/SL spin texture is recorded in the conductance. The scheme also provides a possible detect of the HM/SL spin configuration by transport experiment.

\end{abstract}


\maketitle

\narrowtext

\section{Introduction}

The helimagnet(HM) is a kind of nontrivial magnetic state\cite{Ref13} with its spin spiraling in two or three dimensions characterized by a single spiral wavevector ${\bf{Q}}$.
The skyrmion lattice (SL) is the topologically protected stable spin texture
 with spin vortices forming two-dimensional hexagonal, triangular, or square crystal structures, which was recently discovered in magnetic metal alloys, insulating multiferroic oxides, and the doped semiconductors\cite{SL1, SL2, SL5}. Also, a two-dimensional square lattice of skyrmions was found to be the magnetic ground state of a monolayer
hexagonal Fe film on the Ir(111) surface\cite{SL14}. The SL state was first discovered in the chiral HM\cite{SL1}. The HM and SL can be generalized into a multi-${\bf{Q}}$ description with single ${\bf{Q}}$ in HMs and multiple ${\bf{Q}}$'s in SLs\cite{SL1, SL10, SL11}. These multi-${\bf{Q}}$ states have more complex topological properties and Fourier spectra than earlier-discovered ferromagnet, antiferromagnet, and ferrimagnet. Their spin structures can be detected by neutron scattering\cite{SL1} and Lorentz transmission electron microscopy\cite{SL2}. The Hall effect measurements in the SL metals establish the physics of emergent electrodynamics\cite{SL3, SL4}.

 When a single electron passes through the HM or spin vortex structure, spin-dependent diffraction occurs bearing information of the spin texture. Recently the transport properties of the HM-embedded devices are targeted from different view angles in literature, which lays foundations for the present approach. Manchon \emph{et al.}\cite{Ref42} and us\cite{Ref43} observed the spin-dependent diffraction effect in the transmission at the ferromagnet/HM interface and through the thin-layer HM junction, respectively. Some functional devices were proposed based on the HM such as the persistent spin currents\cite{Ref19}, spin-field-effect transistor\cite{Ref17}, tunneling anisotropic magnetoresistance\cite{Ref18}, and spin resonance\cite{Ref13}. Conductance characteristic of the HM spin configuration and spiral period was found in the Fano resonance spectrum of a quasi-one-dimensional WG containing a thin conducting HM layer as a donor impurity\cite{Ref52}. While the HM-based structures have been extensively investigated particularly for spintronic exploitations\cite{Ref42, Ref43, Ref19, Ref17, Ref18, Ref52, SL12}, scientists are also seeking the potential of the SL in the spintronic arena. Extremely small threshold current to induce spin transfer torque effect was found in bulk SLs\cite{SL6}. Till now, research on the transport properties in SL-based structures is in the preliminary level. There remain many unsolved physical problems. Nevertheless, the SL transport properties are important to reveal its fundamental quantum nature and excavate potential detection methods.

Among various transport devices, we take the SL(HM)-embedded waveguide (WG) as our targeting structure. The most important reason is that the separated levels confined in the WG could avoid phase-randomization by the reservoirs in most other schemes. Resonance and Fano resonance phenomena were theoretically observed in two and quasi-one dimensional WGs containing an impurity point or layer, which correlates propagating and evanescent subbands of the WG\cite{SL7, SL8, SL9, Ref52}. When an electron interacts with the SL vortex, diffraction occurs as a result of the space-dependent spin exchange coupling. Due to the complex geometric configuration, high-order Fourier components should be included in solving the transmissivity. As a primary transport approach, using the WG minimizes the algebra tediousness and does not lose the main characteristics of the SL-related transport.

In this letter, we investigated the transmission and conductance properties in the HM- and SL-film-embedded two-dimensional rectangular electron WG. It is found that under translation of the magnetic film the conductance as a function of the position demonstrates nearly identical pattern to the original spin contour bearing all the spatial symmetry. The scheme provides a possible detect of the HM and SL configuration by transport experiment. The underlying mechanisms are the topology-shaped electron transport.

\section{Theoretical formulation}

The scheme we propose is a two-dimensional rectangular WG containing a thin-film HM/SL layer acting as a scatterer sketched in Fig. 1. The Schr\"{o}dinger equation describing scattering in such a device is
\begin{equation}
\left[ { - \frac{{{\hbar ^2}}}{{2m_e}}{\nabla ^2} + {V_c}\left( {x,y} \right) + {V_{sc}}\left( {x,y,z} \right)} \right]\psi \left( {x,y,z} \right) = E\psi \left( {x,y,z} \right).
\label{eq3}
\end{equation}
The confining potential ${V_c}\left( {x,y} \right) $ of the WG is an infinite
rectangular well with width $L$ and height $H$ across the $x$-$y$ plane. Its eigenstates and corresponding subband energies are
\begin{equation}
{\phi _{mn}}\left( {x,y} \right) = \frac{2}{{\sqrt {HL} }}\sin \frac{{n\pi x}}{L}\sin \frac{{m\pi y}}{H},
\end{equation}
\begin{equation}
{\varepsilon _{mn}} = \frac{{{\hbar ^2}{\pi ^2}}}{{2m_e}}\left( {\frac{{{n^2}}}{{{L^2}}} + \frac{{{m^2}}}{{{H^2}}}} \right),
\end{equation}
with $m$ and $n$ the subband index and $m_e$ the free-electron mass. A rectangular WG is used so that no level degeneracy appears and the third energy level is far above the ground and second energy levels taken into account. We set the incident energy $E$ between the first and second
subbands of the waveguide with the eigenenergies $E_1 =\varepsilon _{11}$, $E_2=\varepsilon _{12}$ and eigenfunctions ${\varphi _1} = {\phi _{11}}$, ${\varphi _2} = {\phi _{12}}$.

The spin vector field ${\bf{n}}\left( {\bf{r}} \right)$ of the HM and SL can be be described in the multi-${\bf{Q}}$ picture\cite{SL1, SL11}.
\begin{equation}
{\bf{n}}\left( {\bf{r}} \right) = \sum\limits_{i = 1}^N {\left[ {{{\bf{n}}_{i1}}\cos \left( {{{\bf{Q}}_i} \cdot {\bf{r}}} \right) + {{\bf{n}}_{i2}}\sin \left( {{{\bf{Q}}_i} \cdot {\bf{r}}} \right)} \right]} ,
\label{eq1}
\end{equation}
with $N$-fold symmetry. In this approach consider $N=1$, $2$, and $3$, which describes a HM, square and triangular SL, respectively. ${{{\bf{Q}}_i}}$ are spiral wave vectors. ${{{\bf{n}}_{i1}}}$ and ${{{\bf{n}}_{i2}}}$ are two unit vectors orthogonal to each other and to ${{{\bf{Q}}_i}}$. All the helices characterized by ${{{\bf{Q}}_i}}$ have the same chirality\cite{SL1} with ${{\bf{Q}}_i} \cdot \left( {{{\bf{n}}_{i1}} \times {{\bf{n}}_{i2}}} \right) = 1$.

We consider four kinds of spin textures with its spin-$z$ component contour depicted in Fig. 2, in which ${{{\bf{Q}}_i}}$, ${{{\bf{n}}_{i1}}}$, and ${{{\bf{n}}_{i2}}}$ can be expressed as follows which does not lose generality.
\begin{equation}
{{\bf{Q}}_i} = q\left\{ {\cos \left[ {\frac{{\left( {i - 1} \right)}}{N}2\pi } \right],\sin \left[ {\frac{{\left( {i - 1} \right)}}{N}2\pi } \right],0} \right\},
\label{eq2}
\end{equation}
\begin{equation}
{{\bf{n}}_{i1}} = \left\{ {\cos \left[ {\frac{{\left( {i - 1} \right)}}{N}2\pi  + \frac{\pi }{2}} \right],\sin \left[ {\frac{{\left( {i - 1} \right)}}{N}2\pi  + \frac{\pi }{2}} \right],0} \right\},
\end{equation}
\begin{equation}
{{\bf{n}}_{i2}} = \left( {0,0,1} \right).
\end{equation}
Here $q = {{2\pi } \mathord{\left/
 {\vphantom {{2\pi } {{a_0}}}} \right.
 \kern-\nulldelimiterspace} {{a_0}}}$ with $a_0$ the spiral period of a single ${\bf{Q}}$ for the multi-${\bf{Q}}$ description of the HM and SLs.
   We investigate the dependence of the transport properties on the HM/SL spin configuration.

The HM/SL thin film is embedded into the WG. We define the total skyrmion number\cite{SL14}
\begin{equation}
S = \frac{1}{{4\pi }}\int_0^H {\int_0^L {{\bf{n}} \cdot \left( {\frac{{\partial {\bf{n}}}}{{\partial x}} \times \frac{{\partial {\bf{n}}}}{{\partial y}}} \right)dxdy} } ,
\label{eq5}
\end{equation}
where ${\bf{n}}$ is the unit vector of the local magnetization defined in Eq. (\ref{eq1}). Different from Ref. \onlinecite{SL14}, the integral is taken over the WG confined area. A complete skyrmion vortex within the WG contributes one topological number and a partial skyrmion vortex contribute a partial topological number depending on the radius and complete whirling rings within the SL. For HMs, the spiral is open without forming closed winding path hence no Berry phase is accumulated and the skyrmion number is zero.

The localized spin-dependent scattering potential of the thin-film HM/SL layer can be approximated by a Dirac-delta function as
\begin{equation}
{V_{sc}}\left( {x,y,z} \right) = \left[ {J{{\bf{n}}}\left( {\bf{r}} \right) \cdot {\bf{\sigma }} + {V_0}} \right]d \delta \left( z \right),
\end{equation}
where $J$ is the sd-type exchange coupling strength between
free electrons and the background spin texture. $V_0$ is the electrostatic potential of the HM/SL. For insulating, semiconducting, and conducting HM/SL materials, $V_0$ is positive and negative correspondingly, which can also be tuned by external electric gates. $d$ is the thickness of the HM/SL layer.

Using the waveguide eigenstates $\left\{ {{\phi _{mn}}} \right\}$ as the expanding
basis of the wave function $\psi \left( {x,y,z} \right)$, Eq. (\ref{eq3}) can be
converted to a set of linear equations. In the matrix formalism,
the equations are\cite{SL7, Ref52}
\begin{equation}
\sum\limits_{\beta n} {\left[ {\left( { - 2iK_{mn}^{\alpha \beta } + V_{mn}^{\alpha \beta }} \right)T_{nl}^{\beta \gamma }} \right]}  =  - 2iK_{ml}^{\alpha \gamma },
\label{eq4}
\end{equation}
where $K_{mn}^{\alpha \beta } = {k_m}{\delta _{mn}} \otimes {I_{\alpha \beta }}$ are elements of the diagonal
matrix of wave vectors with ${k_m} = {{\sqrt {2{m_e}\left( {E - {E_m}} \right)} } \mathord{\left/
 {\vphantom {{\sqrt {2{m_e}\left( {E - {E_m}} \right)} } \hbar }} \right.
 \kern-\nulldelimiterspace} \hbar }$ and is
extended to the wave-spin product space. Also, elements of
the spin-dependent interband transition matrix
\begin{equation}
V_{mn}^{\alpha \beta } = \frac{{2{m_e}}}{{{\hbar ^2}}}\left\langle {\varphi _m^\alpha } \right|\left[ {J{{\bf{n}}}\left( {\bf{r}} \right) \cdot {\bf{\sigma }} + {V_0}} \right]d\left| {\varphi _n^\beta } \right\rangle ,
\end{equation}
where
\begin{equation}
\left| {\varphi _m^u} \right\rangle  = {\varphi _m} \otimes \left( {\begin{array}{*{20}{c}}
1\\
0
\end{array}} \right),\begin{array}{*{20}{c}}
{}&{\left| {\varphi _m^d} \right\rangle  = {\varphi _m} \otimes \left( {\begin{array}{*{20}{c}}
0\\
1
\end{array}} \right)}
\end{array},
\end{equation}
are the waveguide eigenfunctions extended to the spin space in the $\sigma _z$-representation. With both the WG eigenfunctions and the HM/SL spin vector fields (co)sinusoidal functions, $V_{mn}^{\alpha \beta }$ can be analytically obtained.

The spin-dependent transmission
amplitudes $T_{nl}^{\beta \gamma }$ can be obtained from Eq. (\ref{eq4})
by matrix algebra. Here $T_{nl}^{\beta \gamma }$ denotes transmission amplitudes
from the $l$-th subband with spin-$\gamma$ polarization to the $n$-th
subband with spin-$\beta$ polarization. The spin-dependent conductance can be calculated from the two-terminal Landauer formula\cite{SL7}
\begin{equation}
{G^\alpha } = \frac{{{e^2}}}{h}\sum\limits_{mn\beta } {\frac{{{k_m}}}{{{k_n}}}T_{mn}^{*\alpha \beta }T_{mn}^{\alpha \beta }} .
\end{equation}

\section{Numerical results and interpretations}

We consider the transport properties of the HM/SL-embedded two-dimensional rectangular electron WG. We varies the position of the HM/SL film and found that the contour of the conductance as functions of the center coordinates significantly reproduce that of the spin pattern, which promises a potential detection of the HM/SL state. Parameters follow experimental results and those most prominently demonstrating the relation between the conductance and the HM/SL spin configuration are used in the paper and can be extended to a wide value range. The width and height of the WG is set to be $L=120$ nm and $H=60$ nm to demonstrate at least one pattern repeating period within the short width for three-${\bf{Q}}$ SL. The sd-exchange coupling strength $J$ depends on the SL material. Here $J=1.1$ meV is used. We set the incident energy $E=0.18$ meV, which is between the first and second subbands of the waveguide with the eigenenergies $E_1 \approx 0.13$ meV and $E_2 \approx 0.21$ meV. Thickness of the thin-film HM/SL layer $d=20$ nm (following Ref. \onlinecite{SL13}). The electrostatic potential of the HM/SL layer $V_0$ relies on the material and can be tuned by external electric gate potentials. In our numerical treatment $V_0 = - 0.01$ meV meaning a conducting HM/SL shallow well. $a_0 = 19 $ nm (following Ref. \onlinecite{SL1}). Coordinates of the center of the HM/SL are $x_0$ and $y_0$ within the WG.

We consider four types of HM/SL patterns depicted in Fig. 2. Numerical results of the spin-up conductance are shown in Fig. 3. The spin-down conductance demonstrates similar pattern with different absolute values. In the contour as a function of the HM/SL layer center, it could be seen that the conductance significantly reproduce the HM/SL spin configurations with the lattice pattern and symmetry clearly recognizable. The absolute values of the spin-up conductance $G^u$ are close to 1 in unit of $e^2/h$ due to the static shallow well potential $V_0 = - 0.01$ meV used and the pattern resemblance could be most prominently demonstrated. For the three-${\bf{Q}}$ triangular SL, the C2-symmetric rectangular WG squeezes the C3-symmetric triangular lattice at the four sides giving rise to the reshaping of the unit lattice cell in the conductance. The triangular unit cell in the SL spin texture (see Fig. 2) is reshaped into an olive, which also occurs in the total skyrmion number depicted in Fig. 4 and later discussed. We consider the incident electron energy in between the first and second eigenlevels of the WG. The propagating and evanescent modes coherently contribute to the transmission. The third eigenlevel is far above the incident energy and the two levels considered, whose affect can be neglected. This is also justified by normalization of the scattering matrix taking into account the first and second eigenlevels. Work done by the two modes shares some similarity with taking a holographic picture. With single-mode incidence, the eigenlevel of the WG and that of the HM/SL scatterer do not coincide. The HM/SL is complete opaque therefore the conductance is absolute zero. When the evanescent mode enters the transmission, electron transport by two quantum paths interferes. One is direct transport by the propagating level, the other is through scattering to the evanescent mode at the HM/SL scatterer and then back to the propagating level. During interaction with the HM/SL scatter, the spin texture is imprinted in the second path. As evanescent mode cannot transport far, only by interfering with the first path, the spin texture information is recorded. Therefore, the conductance demonstrates a significant resemblance to the HM/SL spin pattern.

Results of the conductance could also be interpreted by the total skyrmion number for the SL case. Contour of the total skyrmion number defined in Eq. (\ref{eq5}) as a function of the SL layer center coordinates is shown in Fig. 4. The integrand
of equation (\ref{eq5}) defines a topological magnetic field that gives rise
to a Lorentz force and the total skyrmion number is proportional to the
accumulated Berry phase in real space of travelling electrons within the total SL layer confined in the WG\cite{SL14}. A complete skyrmion vortex confined within the WG contribute a full single skyrmion number to the total skyrmion number. A partial skyrmion vortex confined within the WG contribute a partial skyrmion number depending on the radius of the confined area. For the three-${\bf{Q}}$ triangular SL, the triagular unit lattice cell of the spin texture is reshaped into an olive in the total skyrmion number pattern due to symmetry compromise, which also occurs in the conductance. As an electron is scattered by the SL quantum well, it feels the Lorentz force of the topological magnetic feel. As the SL spin vortex whirls in the $x$-$y$ plane, the topological magnetic field is in the $z$ direction. The effect of the Lorentz force generated by the magnetic field is to distort wavefunction in the $x$-$y$ plane without changing the transport direction. As the distorted and undistorted wavefunction coming from the two quantum paths interferes, the SL spin texture is recorded. Similar to the SL, the HM has finite topological magnetic field with zero accumulated Berry phase. The wavefunction distortion and path interference occur as well. The conductance records the HM spin pattern as well.

To theoretically explain the experimental observed SL pattern, relative spatial shifts of helices were introduced\cite{SL1} to the spin vector field Eq. (\ref{eq1}). This shift could broaden the skyrmion area and narrow the antiskyrmion area and give rise to identical spin pattern to the experiment. However, in our approach inclusion of the spatial shift of the helices frustrates the analytical solution of $V_{mn}^{\alpha \beta }$ in our theory. In the Supplementary Section S1, we redid our theoretical treatment with phenomenological SL fields close to the experiment pattern. Characteristic periodic curves of the conductance as a function of the SL layer center coordinates are obtained hence the main findings can be reproduced.

\section{Conclusions}

The HM and SL have topologically nontrivial spin texture. The HM and SL can be together generalized into the multi-${\bf{Q}}$ description with the HM a single-${\bf{Q}}$ state, the square SL a double-${\bf{Q}}$ state, and the triangular SL a triple-${\bf{Q}}$ state. We theoretically proposed the device of HM/SL thin film embedded electron WG to investigate the relation between transport properties and the HM/SL spin texture. It is found that by translation of the embedded layer the conductance contour as a function of the HM/SL layer position significantly resemble that of the real-space contour of spin-$z$ component of the HM/SL spin vector field, which has one-to-one correspondence to a particular HM/SL spin configuration. When an electron transports in the WG through the HM/SL scatterer, the two quantum paths, i.e., the propagating mode and through the evanescent mode, interfere. When the electron interacts with the HM/SL scatterer through the second path, the topological magnetic field generalized by the HM/SL spin texture exerts on it and distorts the wavefunction shape in the $x$-$y$ plane. By interference of the two paths, the HM/SL spin texture is recorded in the conductance. The contour pattern of the total skyrmion number as a function of the SL layer center position shares similar pattern with the conductance, which further consolidates the mechanism interpretation.

\section{Acknowledgements}

The author acknowledges enlightening discussions with Wen-Ji Deng, Zhi-Lin Hou, and Li Zhang. This project was supported by the National Natural Science
Foundation of China (No. 11004063) and the Fundamental Research
Funds for the Central Universities, SCUT (No. 2014ZG0044).

\clearpage

\begin{figure}[h]
\includegraphics[height=7cm, width=10cm]{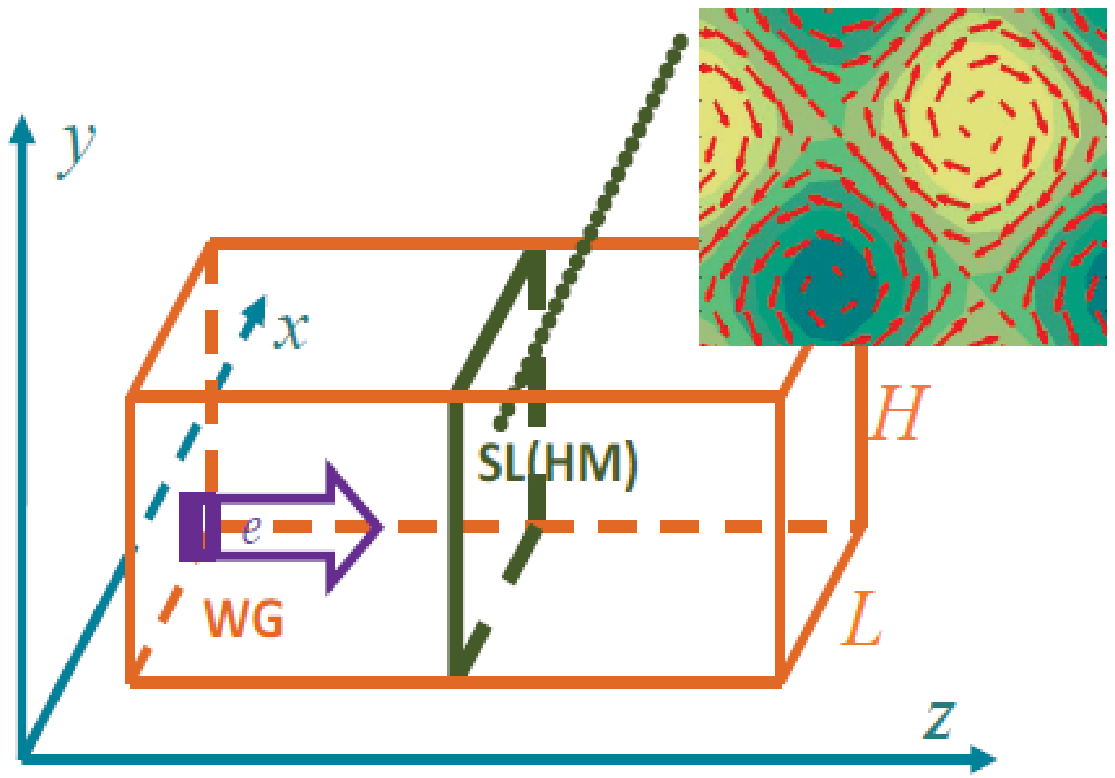}
\caption{Schematics of the SL(HM)-embedded two-dimensional electron WG. The WG confines in the $x$ and $y$ directions with width $L$ and height $H$. Electrons transport in the $z$ direction within the WG. An SL(HM) thin layer is embedded. One type of the spin configuration, the square SL, we considered is amplified in the upper right panel.}
\end{figure}

\clearpage

\begin{figure}[h]
\includegraphics[height=10cm, width=14cm]{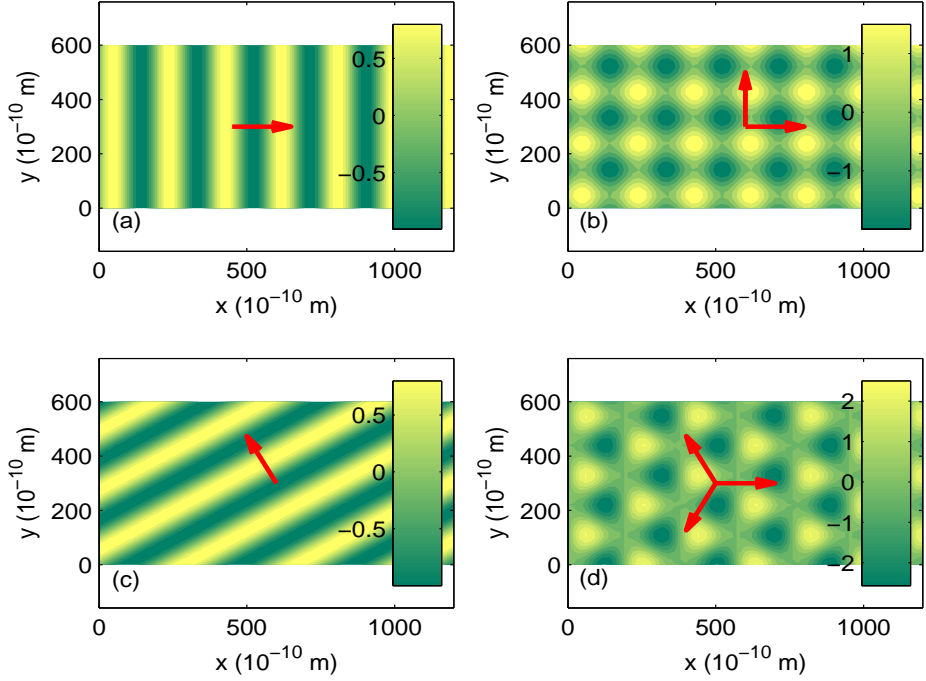}
\caption{The four types of HM/SL spin configurations numerically targeted, which are the HM spiralling in the $x$-direction (a), the two-${\bf{Q}}$ square SL (b), the HM spiralling in the oblique direction with ${\bf{Q}}$ spanned an angle of ${2 \pi}/3$ from the $x$-direction (c), and the three-${\bf{Q}}$ triangular SL (d). The unit spiral wave vectors are the red arrows. The contours are the $z$-components of the HM/SL spin vector field ${\bf{n}}\left( {\bf{r}} \right)$ defined in Eq. (\ref{eq1}). Since all the helices have the same chirality, $z$-component of the vector field also represents the $x$ and $y$ components and characterizes a particular type of HM/SL pattern. The relation between the $z$-component contour and the spin vortex pattern in the $x$-$y$ plane is the same to that shown in the upper right panel of Fig. 1 with skyrmion and antiskyrmion in the green and yellow regions respectively. For $N$-${\bf{Q}}$ HM/SL states the normalization of ${\bf{n}}\left( {\bf{r}} \right)$ is N.  }
\end{figure}

\begin{figure}[h]
\includegraphics[height=10cm, width=14cm]{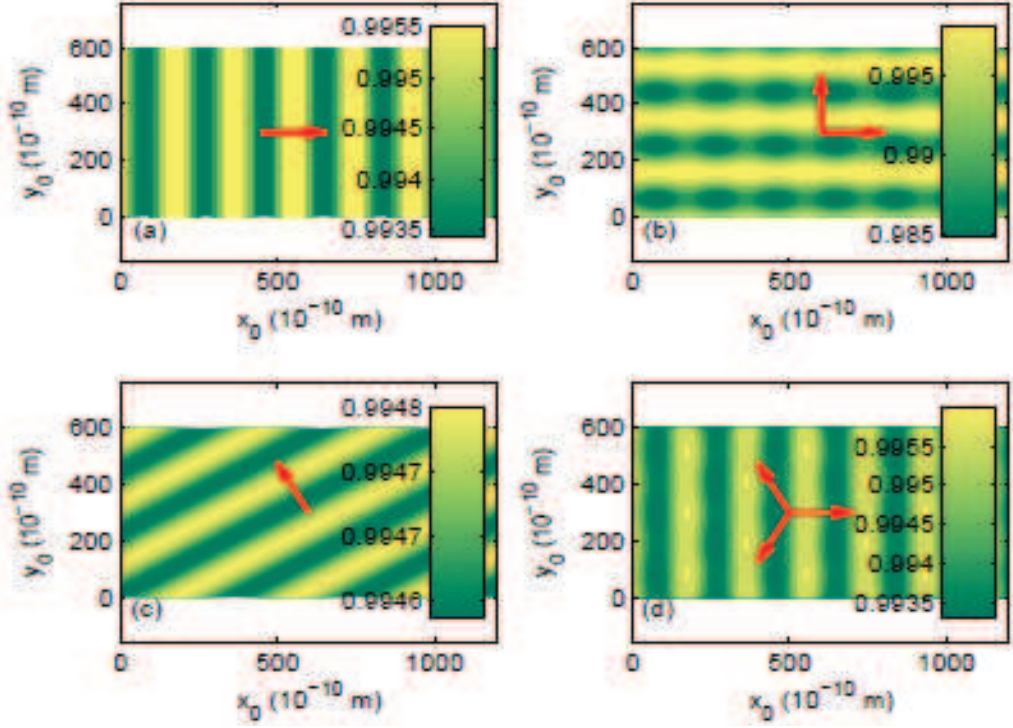}
\caption{Spin-up conductance contour in the unit of ${e^2}/h$ as a function of the center coordinates of the embedded HM/SL layer for the four types of HM/SL spin configurations in correspondence with Fig. 1, which are the HM spiralling in the $x$-direction (a), the two-${\bf{Q}}$ square SL (b), the HM spiralling in the oblique direction with ${\bf{Q}}$ spanned an angle of ${2 \pi}/3$ from the $x$-direction (c), and the three-${\bf{Q}}$ triangular SL (d). Also the unit spiral wave vectors are the red arrows. The spin-down conductance contour demonstrates the same pattern with different absolute values. }
\end{figure}

\begin{figure}[h]
\includegraphics[height=10cm, width=14cm]{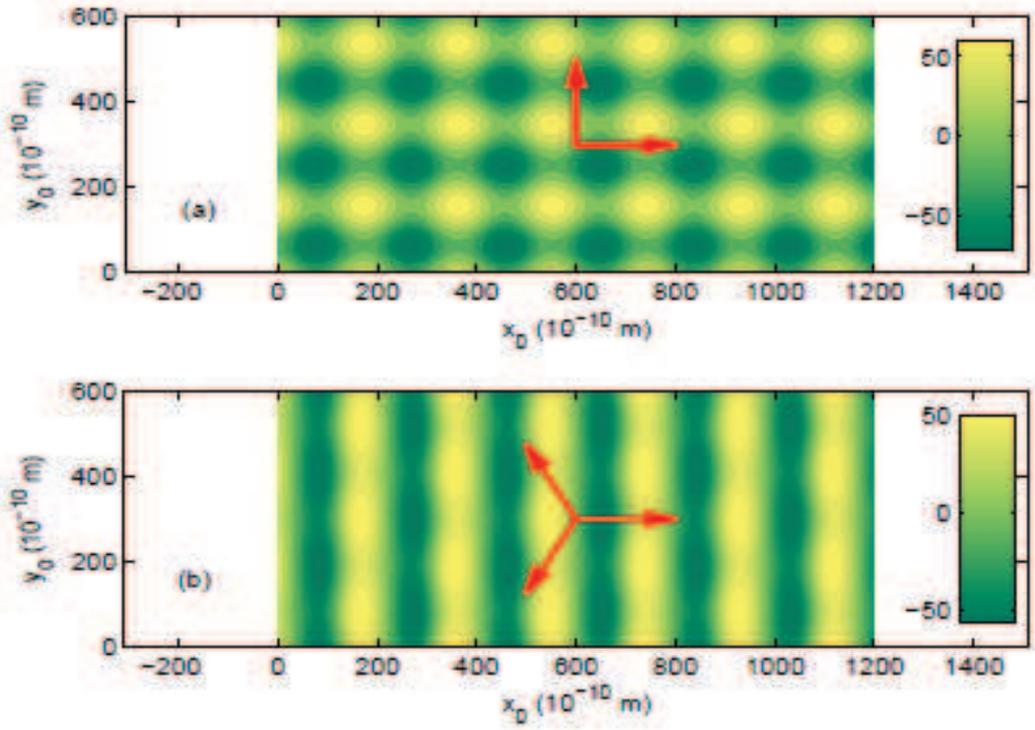}
\caption{Contour of the total skyrmion number defined in Eq. (\ref{eq5}) in the unit of $1/{4 \pi}$ as a function of the center coordinates of the embedded SL layer for the two types of SL pattern in previous figures: (a) the two-${\bf{Q}}$ square SL and (b) the three-${\bf{Q}}$ triangular SL. Also the unit spiral wave vectors are the red arrows. For open single spirals in HMs, the total skyrmion number is zero. }
\end{figure}

\clearpage

\end{document}